\documentclass[12pt, draftclsnofoot, onecolumn]{IEEEtran}

\usepackage{verbatim}
\usepackage[dvips]{graphicx}
\usepackage[cmex10]{amsmath}
\usepackage{amssymb,amsfonts}
\usepackage{multirow}
\usepackage{theorem}
\def\myproof{\noindent{\underline{\textbf{Proof:}}}}

\ifdefined\myproof
\else
\def\myproof{\proof}

\fi

\newtheorem{proof}{Proof}

\usepackage{cite}
\usepackage[breaklinks=true,]{hyperref}
\usepackage{breakurl}

\usepackage[tight,footnotesize]{subfigure}

\usepackage{algorithm}
\usepackage{algorithmic}

\usepackage{subfigure}

\usepackage[justification=centering]{caption}
\usepackage{bbm}

\usepackage{stfloats}
\usepackage{color}
\usepackage{array}
\usepackage{booktabs}

\def\BibTeX{{\rm B\kern-.05em{\sc i\kern-.025em b}\kern-.08em
		T\kern-.1667em\lower.7ex\hbox{E}\kern-.125emX}}

\usepackage{float}
\usepackage{lipsum}
\makeatletter
\newenvironment{breakablealgorithm}
{
	\begin{center}
		\refstepcounter{algorithm}
		\hrule height.8pt depth0pt \kern2pt
		\renewcommand{\caption}[2][\relax]{
			{\raggedright\textbf{\ALG@name~\thealgorithm} ##2\par}%
			\ifx\relax##1\relax 
			\addcontentsline{loa}{algorithm}{\protect\numberline{\thealgorithm}##2}%
			\else 
			\addcontentsline{loa}{algorithm}{\protect\numberline{\thealgorithm}##1}%
			\fi
			\kern2pt\hrule\kern2pt
		}
	}{
		\kern2pt\hrule\relax
	\end{center}
}
\makeatother

\begin{document}
\title{Reinforcement Learning Based Robust Policy Design for Relay and Power Optimization in DF Relaying Networks}

\author{Yuanzhe~Geng,
        Erwu~Liu,~\IEEEmembership{Senior~Member,~IEEE,}
        Rui~Wang,~\IEEEmembership{Senior~Member,~IEEE,}
        Pengcheng~Sun,
        and~Binyu~Lu,~\IEEEmembership{Graduate~Student~Member,~IEEE}
    \thanks{This work is supported in part by the grants from the National Science Foundation of China (No. 42171404, No. 61771345), Science and Technology Commission of Shanghai Municipality (No. 19511102002), and Shanghai Science and Technology Innovation Action Plan Project (No. 21220713100). Corresponding author: Erwu Liu.}

    \thanks{Yuanzhe Geng, Erwu Liu, Pengcheng Sun and Binyu Lu are with the College of Electronic and Information Engineering, Tongji University, Shanghai 201804, China (e-mail: yuanzhegeng@tongji.edu.cn, erwu.liu@ieee.org, pc\_sun2020@tongji.edu.cn, tjabinl@163.com).}

    \thanks{Rui Wang is with the College of Electronic and Information Engineering, Tongji University, Shanghai 201804, China, and also with the Shanghai Institute of Intelligent Science and Technology, Tongji University, Shanghai 201804, China (e-mail: ruiwang@tongji.edu.cn).}
}

\maketitle
\renewcommand\thepage{}

\begin{abstract}
In this paper, we study the outage minimization problem in a decode-and-forward cooperative network with relay uncertainty.
To reduce the outage probability and improve the quality of service, existing researches usually rely on the assumption of both exact instantaneous channel state information (CSI) and environmental uncertainty.
However, it is difficult to obtain perfect instantaneous CSI immediately under practical situations where channel states change rapidly, and the uncertainty in communication environments may not be observed, which makes traditional methods not applicable.
Therefore, we turn to reinforcement learning (RL) methods for solutions, which do not need any prior knowledge of underlying channel or assumptions of environmental uncertainty.
RL method is to learn from the interaction with communication environment, optimize its action policy, and then propose relay selection and power allocation schemes.
We first analyse the robustness of RL action policy by giving the lower bound of the worst-case performance, when RL methods are applied to communication scenarios with environment uncertainty.
Then, we propose a robust algorithm for outage probability minimization based on RL.
Simulation results reveal that compared with traditional RL methods, our approach has better generalization ability and can improve the worst-case performance by about 6\% when evaluated in unseen environments.
\end{abstract}


\section{Introduction}\label{sect_intro}
Cooperative communication has been a hot top in recent years, for its help in realizing resource collaboration between different nodes and obtaining diversified benefits in multi-user scenario.
In cooperative communication system, outage probability is usually employed as a metric to measure the quality-of-service (QoS) of system, which represents the probability that the received signal-to-noise ratio (SNR) falls below a certain threshold \cite{8435942, 9638624}.

In order to improve the QoS of communication system and minimize the outage probability, it is intuitive to optimize the scheme of relay selection and power allocation.
Traditional optimization methods usually rely on the assumption of exact instantaneous channel state information (CSI).
By establishing the probabilistic model where distributions of channel uncertainty are assumed artificially, relay selection and power optimization schemes are then designed following probabilistic analysis \cite{7438886, 9170918, 8673768, 8880523}.
However, it is very possible that perfect instantaneous CSI is difficult to be obtained in situations where channel changes rapidly, and the uncertainties in communication environments may not be observed.
Therefore, traditional methods based on perfect CSI and artificial assumptions may bring estimation bias and mislead the final decisions.

Reinforcement learning (RL) is an emerging technique, and is considered as the third paradigms in machine learning \cite{sutton2018reinforcement}.
The optimization for RL action policy does not need prior knowledge of the environment.
Instead, RL methods employs an agent to interact with the environment, and then learn from the experiences gained during interaction.
Recently, many studies have successfully applied RL technology to help solving resource optimization problems in communications \cite{9072139, 8938826, 9311792}.
In RL-based approaches, the source node usually acts as the agent, which is assumed to be empowered with learning ability.
Therefore, different from traditional methods, we do not need prior knowledge or assumptions about communication environments any more.

In cooperative network, the RL agent will make optimal relay and transmission power decisions directly for the whole system in each time slot, based on its observation of state in previous time slot and the action-reward fed back from the communication system.
In \cite{9137340, 8750861}, Su \textit{et. al} proposed a deep Q network (DQN) based relay scheme, which combines deep neural network (DNN) with the typical Q-learning algorithm, to minimize the outage probability.
Huang \textit{et. al} \cite{9321455} designed a novel decision-assisted deep reinforcement learning (DRL) approach to help relay selection and improve outage performance.
Wang \textit{et. al} \cite{9072416} employed Q-learning method for relay selection to improve energy efficiency.
Similarly, optimization problems of adaptive power control are investigated in \cite{6954557, 7570172, 9018791} and then solved by RL-based methods.
In addition, the joint optimization for relay and power is widely studied with the help of RL technology.
In \cite{9568962, 9120470}, the authors divided relay selection and power optimization into two sub-problems and then solved them through hierarchical RL architecture, in order to maximize total SNR and minimize outage probability.
Huang \textit{et. al} \cite{9448155} proposed asynchronous DRL approaches to maximize system throughput, and Zhang \textit{et. al} \cite{8930580} designed a double DQN architecture to minimize transmission delay.

Despite the successful application and adaptation of RL methods in communication scenarios, there are few literature studies on the robustness of RL based methods.
The studies above are carried out under the condition that elements in the environment do not change, and thus there is a fixed transition dynamic between system states.
In practical situations, however, the location of each relay may be fluid in a cooperative network.
Under such circumstances, the temporal correlation of channel states between each pair of nodes will change, which results in the failure of action policies trained in some environments to generalize to others.
To alleviate the model discrepancy and improve the worst-case performance in changeable environments, some researchers have studied the robustness of action policy in the field of robot control and Atari games \cite{mankowitz2019robust, tessler2019action, janner2019trust, jiang2021monotonic}, which can provide some insights into robust policy analysis in communication scenarios.

In this paper, we analyse the robustness of RL action policy in a changeable decode-and-forward (DF) relaying communication system, where the locations and outage thresholds of each relay are unfixed. Then, we proposed a practicable robust RL algorithm based on proximal policy optimization for relay selection and power allocation, to minimize the outage probability under total power constraints.
Specifically, the contributions of this paper can be summarized as follows.

\begin{itemize}
    \item
    We setup a DF relaying network where uncertainties of both relay location and outage threshold are considered.
    Then, we transform the outage probability minimization problem in uncertain environments into a statistical problem, which can be easily solved by RL methods without any assumptions or prior knowledge of the communication system.

    \item
    We analyse the robustness of RL methods when applied to communication optimization problems, and derive the theorem which gives a lower bound of worst-case performance of action policy in changeable communication environments.

    \item
    Through alternate optimization for both sampling distributions and action policies, we propose an easy-to-implement robust RL algorithm for relay selection and power allocation.
    Simulation results show that the proposed method has good generalization ability, and can improve the worst-case performance when evaluated in unseen environments.

\end{itemize}

The rest of this paper is organized as follows.
Section \ref{sect model} presents the model of our DF relaying system.
Section \ref{sect problem} introduces the variables in RL settings and formulates the outage minimization problem.
Section \ref{sect algorithm} gives the analysis of RL policy robustness in communication scenarios, and then describes our proposed algorithm.
Section \ref{sect evaluation} presents simulation results.
Finally, Section \ref{sect conclusion} concludes this paper.

\section{System Model}\label{sect model}
Considering a typical two-hop cooperative communication network, where a source node $S$ with $n_s$ antennas communicates with a destination node $D$ with $n_d$ antennas via a relay $R_k$.
The relay $R_k$ is selected from a group of feasible relays $\boldsymbol R=\{R_1, R_2, \dots, R_K\}$, each of which has a single antenna and uses decode-and-forward (DF) protocol.

In this network, we assume that the source does not have a direct link to the destination, and the communication from $S$ to $D$ via the selected relay $R_k$ will take two time slots.
In the first time slot, received signal at the selected relay $R_k$ can be represented as
\begin{equation}\label{hop1}
y_{sk}(t)=\sqrt{P_s}\boldsymbol h_{sk}^{\dagger}(t) \boldsymbol x(t) + n_k(t),
\end{equation}
where $P_s \in [0, P_{max}]$ denotes transmission power at source, $\boldsymbol x(t)$ denotes data symbol such that $\|\boldsymbol x(t)\|=1$,
$\boldsymbol h_{sk}(t)$ is a channel vector where each element is a complex Gaussian random variable with zero mean and variance $\sigma^2_{sk} = h_0 d_{sk}^{-\alpha}$, $h_0$ denotes path-loss constant and $\alpha$ denotes path-loss exponent, $d_{sk}$ represents the distance between node $S$ and $R_k$,
$n_k(t)$ denotes the complex Gaussian noise at relay.

In the second time slot, the selected relay forwards the signal to destination, whose received signal can be represented as
\begin{equation}\label{hop2}
\boldsymbol y_{kd}(t)=\sqrt{P_r} \boldsymbol h_{kd}^{\dagger}(t) y_{sk}(t) + \boldsymbol n_d(t),
\end{equation}
where $P_r \in [0, P_{max}]$ denotes transmission power at relay, $\boldsymbol h_{kd}(t)$ denotes channel vector between $R_k$ and $D$, and similarly, each of its elements is a complex Gaussian random variable with zero mean and variance $\sigma^2_{kd} = h_0 d_{kd}^{-\alpha}$, $\boldsymbol n_d(t)$ denotes complex Gaussian noise at destination.

Then, we have the expressions for mutual information as follows \cite{4107949, 8532123}.
\begin{equation}\label{mi_r} \begin{aligned}
I_k=\log_2(1+P_s {\|\boldsymbol h_{sk}\|}^2/ P_0), I_d=\log_2(1+P_r {\|\boldsymbol h_{kd}\|}^2/ P_0),
\end{aligned}\end{equation}
where we assume $P_0$  to be the noise power at both relay and destination for simplicity.

\section{Problem Formulation}\label{sect problem}
In situations where channel states change quickly, it is usually difficult to obtain instantaneous CSI immediately, and delay often occurs.
For such scenario, traditional resource allocation methods are not applicable, and thus we turn to RL methods for solutions.
Suppose that the source node is empowered with learning ability, and acts as the RL agent in our communication environment.
The agent determines the optimal relay and power for current moment, based on its observation of the channel state in previous time slot.
After performing relay selection and power allocation, it receives a binary reward from environment, which indicates whether the action leads to a successful communication under channel state in current time slot.
In this section, we model this process as an Markov decision process (MDP), and give the definition of MDP variables according to our cooperative communication scenario.
Then, we formulate the outage minimization problem based on these settings.

\subsection{Environment Parameter Space}\label{subsect_env}
Different from the settings in traditional RL-based studies, we further consider the influence of environmental uncertainty on the interaction between RL agent and communication system, and introduce the concept of environment parameter.
Environment parameters, which can also be called as 'context' \cite{kirk2021survey}, are a class of variables in environment that can affect the state transition dynamics of the system and the feedback reward.

In our problem, we assume that the locations of relays change at regular intervals, and each of relays may appear at a set of $L$ possible locations.
Therefore, distances from the $k$-th relay to source and destination are also not fixed, which can be represented as $d_{sk} \in \boldsymbol d_{sk} = \{ d_{sk}^1, d_{sk}^2, \dots, d_{sk}^L \}$ and $d_{kd} \in \boldsymbol d_{kd} = \{ d_{kd}^1, d_{kd}^2, \dots, d_{kd}^L \}$, respectively.
Besides, the rate threshold $\lambda_k$ that the $k$-th relay needs to satisfy when decoding the received signal can also be assumed to be unfixed, which can be represented as $\lambda_k \in \boldsymbol\lambda_{k} = \{ \lambda_{k}^1, \lambda_{k}^2, \dots, \lambda_{k}^{J} \}$.
Then, full environment parameter space can be denoted as
\begin{equation}\label{eq_env} \mathcal{P}_t \triangleq [d_{sk}(t), d_{kd}(t), \lambda_{k}(t)], \end{equation}
where integer $k \in [1, K]$.
The first and the second parameter denote the distance from each relay to source and destination in time slot $t$, respectively.
The third parameter denotes rate threshold requirements of each relay.

Therefore, the realization of these three parameters form a group, and then construct a unique communication environment.
When at least one parameter of them changes, the environment with which the RL agent interacts changes accordingly.

\subsection{State Space}\label{subsect_state}
In scenarios where wireless channels change rapidly, it is difficult to obtain instantaneous CSI.
Therefore, in our system, the state space for RL agent is the observation of channel states between any two nodes in the previous time slot, which can be represented as
\begin{equation}\label{eq_state} \mathcal{S}_t \triangleq [\boldsymbol h_{sk}(t-1), \boldsymbol h_{kd}(t-1)], \end{equation}
where integer $k \in [1, K]$.

In order to characterize the temporal correlation between time slots for each channel, we assume that wireless channels of any link in our system all follow the Markov model.
In terms of each possible location of relay $R_k$, we sample $M$ channel states for its $s-r$ link which can be simply written as $\{ state^{sk}_1, state^{sk}_2, \dots, state^{sk}_M \}$.
Then, under the same environment parameter $p_t=p_{t-1}=d_{sk}$, the transition probability $p(\cdot)$ from the $i$-th state to the $j$-th state can be represented as
\begin{equation}\label{eq_state_2} p_{ij}(s_t|s_{t-1}, d_{sk})={\rm Prob}(s_t=state^{sk}_j | s_{t-1}=state^{sk}_i, d_{sk}). \end{equation}
Note that, the transition probabilities will change according to current environment parameters.
Operations and definitions for $r-d$ link are similar.

\subsection{Action Space}\label{subsect_action}
The source node is considered as the agent in our RL settings.
It needs to select one relay, and make power allocation for both source and the relay.
Considering that the total power for source and relay is limited to $P_{max}$, we can assume the sum transmission power for these two nodes is just $P_{max}$ and then equivalently replace $P_r$ by $P_{max} - P_s$.
Then, full action space can be represented as
\begin{equation}\label{eq_action} \mathcal{A}_t \triangleq [a^{R}(t), a^{P_s}(t)], \end{equation}
where 
\begin{equation}\label{eq_condition_1} a^{R}(t) \in \{1, 2, \dots, K\} \end{equation}
and
\begin{equation}\label{eq_condition_2} a^{P_s}(t) \in [0, P_{max}]. \end{equation}

\subsection{Reward Function and Optimization Problem}\label{subsect_action}
In DF relaying system, the outage probability minimizing problem that jointly optimizes relay selection and power allocation can be intuitively formulated as follows.
\begin{equation}\label{eq_problem_1_1}
    \min\limits_{\mathcal{A}_t} \ {\rm Prob}(I_k<\lambda_k, I_d<\lambda_d),
\end{equation}
where the positive scalars $\lambda_k$ and $\lambda_d$ are denoted to be the outage threshold for the $k$-th relay and destination.

Based on (\ref{eq_problem_1_1}), we consider a binary reward case that RL agent can only know whether current communication is successful or not after it performs actions.
Then, by using the indicator function, our reward function can be designed as
\begin{equation}\label{eq_reward}
    f(a^{R}, a^{P_s}; \boldsymbol h, \boldsymbol p) \triangleq \mathbbm{1}_{I<\lambda}=\left\{
        \begin{aligned}
            &1, \ {\rm if}\ I_k<\lambda_k, I_d<\lambda_d \ \\
            &0, \ \ \ \ \ \ {\rm otherwise}
        \end{aligned}
    \right.
\end{equation}
Since the expectation of an indicator function can be used to calculate the probability of the an event, when the indicator function is employed to represent each occurrence of the origin event.
Therefore, we can reformulate the optimization problem for minimizing outage probability of our communication system as follows.
\begin{equation}\label{eq_problem_1_2} \begin{aligned}
\min\limits_{\mathcal{A}_t} \ \mathbb{E} \left[ \frac{1}{T}\sum\limits_{t=1}^T f\big(a^{R}(t), a^{P_s}(t);\boldsymbol h(t), \boldsymbol p(t) \big) \right] \ \
s.t. \ (\ref{eq_condition_1}),(\ref{eq_condition_2}),
\end{aligned}\end{equation}
which then can be solved by using RL methods.

\section{Robust Proximal Policy Based Solution}\label{sect algorithm}
To improve the outage probability in changeable environments with uncertainty, we derive the theorem that gives the lower bound of RL action policy's worst-case performance in communication scenarios.
On this basis, we further propose a RL policy optimization algorithm for relay selection and power allocation, which is easy to implement in practical use.

\subsection{Robust RL Policy Analysis for Communication Scenarios}\label{subsect_theorem}
\textbf{Lemma 1. } \textit{Suppose two joint distributions $p_1(x,y)=p_1(x)p_1(y|x)$ and $p_2(x,y)=p_2(x)p_2(y|x)$, the total variation distance of the joint can be bounded by:}
\begin{equation}\label{lemma_1}\begin{aligned}
 D_{TV}\big(p_1(x,y)||p_2(x,y)\big) \leq \mathbb{E}_{x \sim p_1}\big[D_{TV}\big(p_1(y|x)||p_2(y|x)\big)\big] + D_{TV}\big(p_1(x)||p_2(x)\big).
\end{aligned}\end{equation}

\textbf{Lemma 2. } \textit{Suppose the state distributions $p_1(s_1)$ and $p_2(s_1)$ in initial time slot are the same, the total variation distance in the state marginal can be bounded by:}
 \begin{equation}\label{lemma_2}\begin{aligned}
 D_{TV}\big(p_1(s_t)||p_2(s_t)\big) \leq (t-1)\max_t \mathbb{E}_{s_{t-1} \sim p_1} D_{TV}\big(p_1(s_t|s_{t-1})||p_2(s_t|s_{t-1}) \big).
\end{aligned}\end{equation}

Based on Lemma 1 and Lemma 2, we have the following lower bound expression of worst-case performance under the guide of RL action policy in communication scenarios.

\textbf{Theorem 1. }
\textit{Suppose that the expected total variance distance between two policies is bounded by $\epsilon(\tilde{\pi}||\pi)$, and that between two transition dynamics is bounded by $\epsilon(p_w||p)$.
Then in MDPs that reward is bounded by $r_{max}$, by updating RL agent's policy $\pi$ to a new one $\tilde{\pi}$, we have the following bound: }
\begin{equation}\label{theorem_1}
\eta(\tilde{\pi}|p_w) \geq \mathbb{E}_{p \sim P}[\eta(\tilde{\pi}|p)] -\frac{4r_{max}\gamma}{1-\gamma}\epsilon(\tilde{\pi}||\pi) - \frac{2r_{max}\gamma^2}{(1-\gamma)^2}\mathbb{E}_{p \sim P}[\epsilon(p_w||p)],
\end{equation}
where $p$ denotes environment parameter that samples according to a distribution $P$ over the whole environment parameter space $\mathcal{P}$, and $p_w$ denotes the worst environment parameter that generates the worst-case performance under current policy.
$\epsilon(\tilde{\pi}||\pi) = \max_t \mathbb{E}_{s_t \sim p(\cdot|p_w)} D_{TV}\big(\tilde{\pi}(a_t|s_t) || \pi(a_t|s_t) \big)$,
and $\epsilon(p_w||p) = \max_{t} \mathbb{E}_{s_{t-1} \sim p(\cdot|p_w)} D_{TV}\big( p(s_t|s_{t-1}, p_w) || p(s_t|s_{t-1}, p) \big)$.

\textbf{Proof. }To obtain the lower bound of worst-case performance in Theorem 1, we first construct the following two parts.
\begin{equation}\label{proof_all}\begin{aligned}
\eta(\tilde{\pi}|p_w) - \mathbb{E}_{p \sim P}[\eta(\tilde{\pi}|p)]
&=\Big( \eta(\tilde{\pi}|p_w) - \eta(\pi|p_w) \Big) + \Big( \eta(\pi|p_w) - \mathbb{E}_{p \sim P}[\eta(\tilde{\pi}|p)]  \Big)
\end{aligned}\end{equation}

In terms of the first part, by using Lemma 1, we have
\begin{equation}\label{proof_1_1}\begin{aligned}
&\big| \eta(\tilde{\pi}|p_w) - \eta(\pi|p_w) \big| \\
&= \sum\limits_t \gamma^t \big| \sum\limits_{s_t,a_t} \big( p(s_t,a_t|p_w,\tilde{\pi}) - p(s_t,a_t|p_w,\pi) \big) r(s_t,a_t) \big| \\
&\leq r_{max} \sum\limits_t \gamma^t \sum\limits_{s_t,a_t} \big| p(s_t,a_t|p_w,\tilde{\pi}) - p(s_t,a_t|p_w,\pi) \big| \\
&= 2r_{max} \sum\limits_t \gamma^t D_{TV}\big( p(s_t,a_t|p_w,\tilde{\pi}) || p(s_t,a_t|p_w,\pi) \big) \\
&\leq 2r_{max} \sum\limits_t \gamma^t \Big(
  \mathbb{E}_{s_t \sim p(\cdot|p_w,\tilde{\pi})} D_{TV}\big(\tilde{\pi}(a_t|s_t) || \pi(a_t|s_t) \big) + D_{TV}\big(p(s_t|p_w,\tilde{\pi}) || p(s_t|p_w,\pi) \big)
\Big).
\end{aligned}\end{equation}
Considering that in communication scenarios, the probability of the occurrence of channel state is unrelated to RL agent's action policy.
As a result, the latter total variation (TV) distance term is equal to 0, and thus
\begin{equation}\label{proof_1_2}\begin{aligned}
\big| \eta(\tilde{\pi}|p_w) - \eta(\pi|p_w) \big|
&\leq 2r_{max} \sum\limits_t \gamma^t \Big( \mathbb{E}_{s_t \sim p(\cdot|p_w)} D_{TV}\big(\tilde{\pi}(a_t|s_t) || \pi(a_t|s_t) \big) \Big).
\end{aligned}\end{equation}
With the assumption that $\epsilon(\tilde{\pi}||\pi) = \max_t \mathbb{E}_{s_t \sim p(\cdot|p_w)} D_{TV}\big(\tilde{\pi}(a_t|s_t) || \pi(a_t|s_t) \big)$, we have
\begin{equation}\label{proof_1_3}\begin{aligned}
\eta(\tilde{\pi}|p_w) - \eta(\pi|p_w) \geq -2r_{max} \sum\limits_t \gamma^t \epsilon(\tilde{\pi}||\pi)
\end{aligned}\end{equation}

In terms of the second part, by using Lemma 1 and Lemma 2, we have
\begin{equation}\label{proof_2_1}\begin{aligned}
&\big| \eta(\pi|p_w) - \mathbb{E}_{p \sim P}[\eta(\tilde{\pi}|p)] \big| \\
&\leq \mathbb{E}_{p \sim P} \big| \eta(\pi|p_w)-\eta(\tilde{\pi}|p) \big| \\
&= \mathbb{E}_{p \sim P} \big| \sum\limits_t \gamma^t \sum\limits_{s_t,a_t} \big( p(s_t,a_t|p_w,\pi) - p(s_t,a_t|p,\tilde{\pi}) \big) r(s_t,a_t) \big| \\
&\leq r_{max} \mathbb{E}_{p \sim P} \sum\limits_t \gamma^t \sum\limits_{s_t,a_t} \big|  p(s_t,a_t|p_w,\pi) - p(s_t,a_t|p,\tilde{\pi})  \big| \\
&= 2r_{max} \sum\limits_t \gamma^t \mathbb{E}_{p \sim P}[D_{TV}(p(s_t,a_t|p_w,\pi) || p(s_t,a_t|p,\tilde{\pi}))] \\
&\leq 2r_{max} \sum\limits_t \gamma^t \mathbb{E}_{p \sim P} \big[
  \mathbb{E}_{s_t \sim p(\cdot|p_w,\pi)}D_{TV}\big(\pi(a_t|s_t)||\tilde{\pi}(a_t|s_t) \big) + D_{TV}\big(p(s_t|p_w,\pi)||p(s_t|p,\tilde{\pi}) \big)
\big] \\
&= 2r_{max} \sum\limits_t \gamma^t \big[
  \mathbb{E}_{s_t \sim p(\cdot|p_w)}D_{TV}\big(\pi(a_t|s_t)||\tilde{\pi}(a_t|s_t) \big) + \mathbb{E}_{p \sim P} D_{TV}\big(p(s_t|p_w)||p(s_t|p) \big)
\big] \\
&\leq 2r_{max} \sum\limits_t \gamma^t \big[
  \epsilon(\tilde{\pi}||\pi) + \mathbb{E}_{p \sim P} D_{TV}\big(p(s_t|p_w)||p(s_t|p))
\big] \\
&\leq 2r_{max} \sum\limits_t \gamma^t \big[
  \epsilon(\tilde{\pi}||\pi) + (t-1) \mathbb{E}_{p \sim P} \max_{t} \mathbb{E}_{s_{t-1} \sim p(\cdot|p_w)} D_{TV}\big( p(s_t|s_{t-1}, p_w) || p(s_t|s_{t-1}, p)  \big)
\big].
\end{aligned}\end{equation}
With the assumption that $\epsilon(p_w||p) = \max_{t} \mathbb{E}_{s_{t-1} \sim p(\cdot|p_w)} D_{TV}\big( p(s_t|s_{t-1}, p_w) || p(s_t|s_{t-1}, p) \big)$, we have
\begin{equation}\label{proof_2_2}\begin{aligned}
\eta(\pi|p_w) - \mathbb{E}_{p \sim P}[\eta(\tilde{\pi}|p)] \geq -2r_{max} \sum\limits_t \gamma^t \big(
  \epsilon(\tilde{\pi}||\pi) + (t-1) \mathbb{E}_{p \sim P}[\epsilon(p_w||p)]
\big).
\end{aligned}\end{equation}

Then, by combining the two parts in (\ref{proof_1_3}) and (\ref{proof_2_2}), we can obtain the following result.
\begin{equation}\label{proof_0}\begin{aligned}
&\eta(\tilde{\pi}|p_w) - \mathbb{E}_{p \sim P}[\eta(\tilde{\pi}|p)] \\
&\geq -4r_{max} \sum\limits_t \gamma^t \epsilon(\tilde{\pi}||\pi) - 2r_{max} \sum\limits_t \gamma^t (t-1) \mathbb{E}_{p \sim P}[\epsilon(p_w||p)] \\
&=-\frac{4r_{max}\gamma(1-\gamma^t)}{1-\gamma}\epsilon(\tilde{\pi}||\pi) - \frac{2r_{max}\big[\gamma^2-(t-1)(1-\gamma)\gamma^{t+1}-\gamma^{t+2}]}{(1-\gamma)^2}\mathbb{E}_{p \sim P}[\epsilon(p_w||p)] \\
&> -\frac{4r_{max}\gamma}{1-\gamma}\epsilon(\tilde{\pi}||\pi) - \frac{2r_{max}\gamma^2}{(1-\gamma)^2}\mathbb{E}_{p \sim P}[\epsilon(p_w||p)].
\end{aligned}\end{equation}
By moving the expectation term on the left side of the inequality to the right, we can get the result in (\ref{theorem_1}).
$\hfill \blacksquare$

\subsection{Practicable RL Algorithm Design}\label{subsect_algorithm}
Theorem 1 provides a lower bound for the performance of action policy under the worst environment parameter case.
In order to improve the worst-case performance, we can equivalently maximize the right side in (\ref{theorem_1}), that is,
\begin{equation}\label{problem_2}\begin{aligned}
\max_{\tilde{\pi}, P} \Big[\mathbb{E}_{p \sim P}[\eta(\tilde{\pi}|p)] -\frac{4r_{max}\gamma}{1-\gamma}\epsilon(\tilde{\pi}||\pi) - \frac{2r_{max}\gamma^2}{(1-\gamma)^2}\mathbb{E}_{p \sim P}[\epsilon(p_w||p)] \Big],
\end{aligned}\end{equation}
where the first term denotes the average performance over all environment parameters, and the second and the third term can be regarded as TV distance penalization.

In terms of optimization problem in (\ref{problem_2}), we take the two-step alternate optimization method to reduce the amount of computation. Firstly, with parameter $\tilde{\pi}$ fixed, the parameter $P$ will be optimized such that
\begin{equation}\label{problem_2_step1}\begin{aligned}
\max_{P} \Big[\mathbb{E}_{p \sim P}[\eta(\tilde{\pi}|p)] - \frac{2r_{max}\gamma^2}{(1-\gamma)^2}\mathbb{E}_{p \sim P}[\epsilon(p_w||p)] \Big],
\end{aligned}\end{equation}
where the term including policy divergence $\epsilon(\tilde{\pi}||\pi)$ is omitted as it is unrelated to $P$.
Considering that the objective function is linear to optimization parameter $P$, this task can be operated by assigning higher probabilities to better environment parameters.
Then, the corresponding experience samples of these environment parameters, which can produce better performance in objective function (\ref{problem_2_step1}) than the worst parameter does, will be selected for policy training.

However, in practical situations, it is unfortunate for model-free RL methods that the transition dynamic of channel states is unavailable, and thus the value of $\epsilon(p_w||p)$ cannot be estimated.
Considering that transition dynamics will vary with environment parameters, we follow the empirical method proposed in \cite{jiang2021monotonic} that the TV distance between environment parameters is used to replace the TV distance between transition dynamics, to solve this issue.
As a result, the objective in (\ref{problem_2_step1}) can be alternatively written as $\max_{P} \mathbb{E}_{p \sim P} \big[\eta(\tilde{\pi}|p) - \zeta D_{TV}(p_w||p) \big]$, where $\zeta$ is used to denote the value of constant part.

Secondly, with parameter $P$ fixed, the parameter $\tilde{\pi}$ will be optimized such that
\begin{equation}\label{problem_2_step2}\begin{aligned}
\max_{\tilde{\pi}} \Big[\mathbb{E}_{p \sim P}[\eta(\tilde{\pi}|p)] - \frac{4r_{max}\gamma}{1-\gamma}\epsilon(\tilde{\pi}||\pi) \Big],
\end{aligned}\end{equation}
where the term including transition dynamic divergence $\epsilon(p_w||p)$ is omitted as it is unrelated to $\tilde{\pi}$.
This task belongs to the typical trust region policy optimization problems in RL \cite{schulman2015trust, schulman2017proximal}, which can be solved by proximal policy optimization (PPO) algorithms.

\begin{figure}[htb]
	\centering
	\includegraphics[scale=0.85]{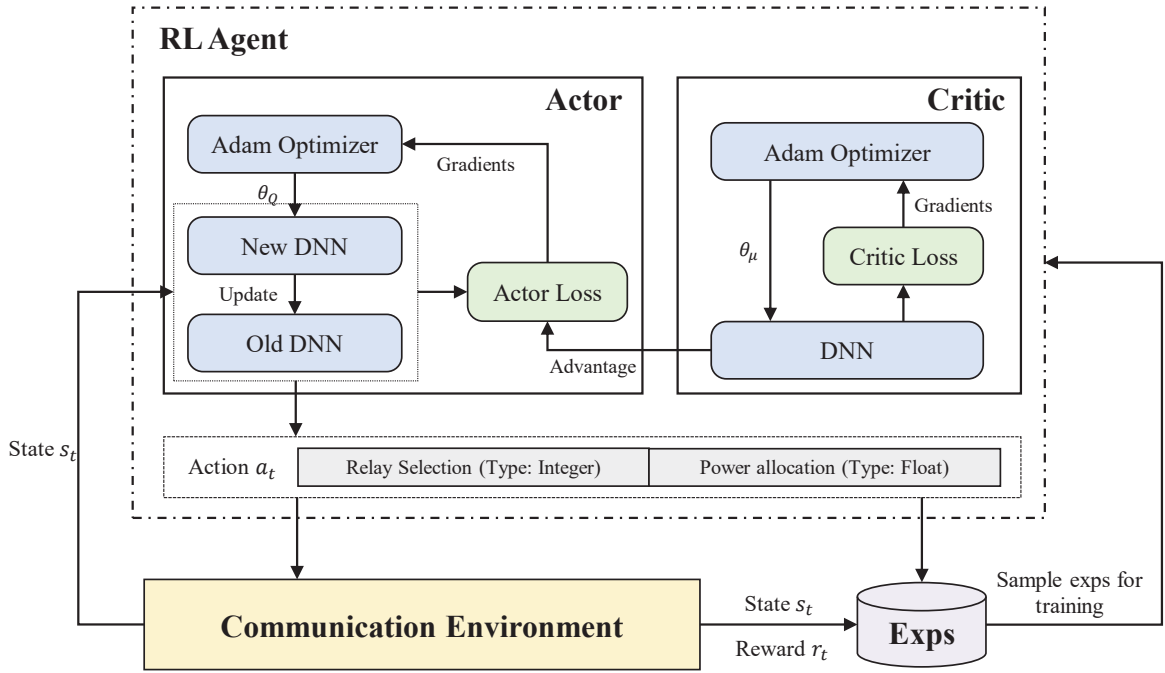}
    \caption{Actor-critic framework for relay and power optimization.}
	\label{DRL_structure}
\end{figure}

More specifically, we use the actor-critic structure to implement PPO, as shown in Fig. \ref{DRL_structure}.
The critic estimates action-value by employing a DNN with parameter $\theta_Q$, and similar to other value-based RL methods, it tries to minimize the following loss function.
\begin{equation}\label{critic_loss}
L(\theta_Q) = \delta_t^2(s_t, a_t; \theta_Q),
\end{equation}
where $\delta_t(s_t, a_t; \theta_Q) = r_t + \gamma \max\nolimits_{a_{t+1}}Q(s_{t+1}, a_{t+1}; \theta_Q) - Q(s_t, a_t; \theta_Q)$ is called advantage (also known as temporal difference error).
Then, the network parameter will be updated by using Adam optimizer.

The actor uses a DNN with parameter $\theta_\mu$ to learn from experience samples, and perform actions for both relay selection and power allocation.
Note that, the actor also keeps its old network parameter $\theta_\mu^-$ before last parameter update.
Then, the following clipped surrogate loss function will be calculated through both new and old actor network \cite{schulman2017proximal}.
\begin{equation}\label{actor_loss}
L(\theta_\mu) = \min \Big[ \frac{\tilde{\pi}(a_t|s_t;\theta_\mu)}{\pi(a_t|s_t;\theta_\mu^-)} \delta_t(s_t, a_t; \theta_Q), {\rm clip} \big( \frac{\tilde{\pi}(a_t|s_t;\theta_\mu)}{\pi(a_t|s_t;\theta_\mu^-)}, 1-\kappa, 1+\kappa \big) \delta_t(s_t, a_t; \theta_Q) \Big],
\end{equation}
where $\kappa$ is a hyper-parameter for clipping the probability ratio.
Similarly, to minimize the loss function above, Adam optimizer is used when updating actor network parameter.

Due to the characteristics of the actor network, both variables representing relay and power are considered to be continuous first.
However, our outage probability minimization problem in (\ref{eq_problem_1_2}) is a Mixed-Integer Non-Linear Programming problem, which is constrained by discrete region in relay selection and continuous region in power allocation.
Therefore, in order to satisfy the constraints of optimization variables, RL agent will perform the Floor operation on the result of relay selection outputted by the actor network, while directly adopt the outputted power.

The two steps of optimization above will be performed alternatively in each training episode.
Pseudocode of our proposed algorithm can be found in Algorithm \ref{RPPO}.

\begin{breakablealgorithm}
    \caption{Robust Proximal Policy for Relay Selection and Power Allocation}
    \label{RPPO}
    \begin{algorithmic}[1]
        \STATE Initialize network parameter $\theta_Q$ for critic.
        \STATE Initialize network parameter $\theta_\mu$ for actor, and $\theta_\mu^- = \theta_\mu$ for old actor.

        \FOR{episode $u=1,2,\dots,u_{max}$}
            \STATE Sample a batch of environment parameters $\{p_1, p_2, \dots, p_{m_{max}}\}$.
            \FOR {parameter index $m=1,2,\dots, m_{max}$}
                \STATE Reset the environment by parameter $p_m$, and get the initial state.
                \FOR{trial index under current environment $l=1,2,\dots,l_{max}$}
                    \FOR{time slot $t=1,2,\dots,t_{max}$}
                        \STATE Choose action $a_t=\tilde{\pi}(s_t;\theta_\mu)$ to determine the relay and transmission power.
                        \STATE Execute action $a_t$, and observe reward $r_{t}$ and next state $s_{t+1}$.
                        \STATE Collect and save current experience sample.
                        \STATE Update current state.
                    \ENDFOR
                \ENDFOR
                \STATE Calculate the average cumulative discounted reward according to $\eta(\tilde{\pi}|p_m)=\frac{1}{l_{max}} \sum\nolimits_{l}\sum\nolimits_{t}\gamma^t r_t$.
            \ENDFOR
            \STATE Find the worst environment parameter according to $p_w=\arg\min\nolimits_{p_m}\eta(\tilde{\pi}|p_m)$.
            \FOR {parameter index $m=1,2,\dots, m_{max}$}
                \IF {$\eta(\tilde{\pi}|p_m) - \zeta D_{TV}(p_w||p_m) \geq \eta(\tilde{\pi}|p_w)$}
                    \STATE Save current actor network parameter by $\theta_\mu^- = \theta_\mu$.
                    \STATE Calculate loss functions in (\ref{critic_loss}) and (\ref{actor_loss}) through experience samples obtained under environment parameter $p_m$, and minimize the loss functions.
                    \STATE Update network parameters for critic and actor  through Adam optimizer.
                \ENDIF
            \ENDFOR
        \ENDFOR
    \end{algorithmic}
\end{breakablealgorithm}

\section{Evaluation}\label{sect evaluation}
In this section, we first introduce the settings of environment that the RL agent interacts with.
Then, we carry out several experiments to evaluate the performance of our proposed method.

\subsection{Experiment Setup}\label{subsect_settings}
In the two-hop cooperative relay network, we assume that the source and destination are fixed, each mobile relay has $L=5$ possible locations.
The outage rate threshold for destination is fixed to be $\lambda_d=1.0$, and that for each mobile relay will be randomly selected from $J=5$ possible thresholds.
The maximum total power for transmission $P_{max}$, which is the sum of power for source and relay, is limited to 0.1W.

To implement our proposed algorithm, we use DNN architecture with six layers for both actor and critic network.
The DNN for critic network consists of one input layer, one output layer, and four hidden layers.
The number of neurons in input layer is equal to the numbers of states, and the number of neurons in output layer is equal to 1, which outputs the action-value estimation result.
Each hidden layer has 20 neurons, and Tanh function is employed to act as the activation function.
The structure of DNN for actor network is similar to that in critic network.
The only difference is that, its output layer consists of two parts, which outputs mean and standard deviation of a normal distribution respectively.
Then the distribution guides RL agent to perform actions with probabilities.
Besides, the learning rates in Adam optimizer for actor and critic network are set to be 0.001 and 0.005, respectively.

For comparison, we use the following methods as baselines in our experiments.

\textbf{Original PPO (PPO)}:
PPO is a famous on-policy methods, which has been taken as the benchmark by many works, due to its trust region design for stable policy update.
Consider that our proposed algorithm is designed based on PPO, we can therefore use its original version as a baseline.

\textbf{Deep Deterministic Policy Gradient (DDPG)}:
DDPG is an off-policy method that combines value estimation method and policy gradient method together.
It is also applicable for optimization problems with continuous action space.
We will take DDPG for comparison, and RL settings for it are the same with those for our proposed method.

\textbf{Deep Q Network (DQN)}:
DQN is a typical off-policy method which has been widely used in the field of communication for resource optimization problems.
Note that, DQN can only solve problems with discrete action spaces.
Therefore, when employing this method, we need to divide the transmission power into several levels for RL agent to choose from.

\textbf{Random Selection (Random)}:
For each time slot, the RL agent will randomly select a relay to perform cooperative communication with random transmission power.

\subsection{Numerical Results}\label{subsect_results}
In the first experiment, we only choose the location of relays as the variable in environment, and evaluate the performance of different methods in situations that transition dynamics of system states keep varying.
Although we set $L=5$ possible locations for each relay, we only select three possible locations for each relay and use them for training.
It means that the other two possible locations will be unseen situations to the RL agent.
Then, all possible locations will be considered in testing process, in order to evaluate the generalization ability of different methods.

\begin{figure*}[hbt]
	\centering

    \vspace{-0.35cm} 
	\subfigtopskip=1pt 
	\subfigbottomskip=1pt 
	\subfigcapskip=-4pt 

    \subfigure[ ]{
		\label{avg_train1}
		\includegraphics[scale=0.5]{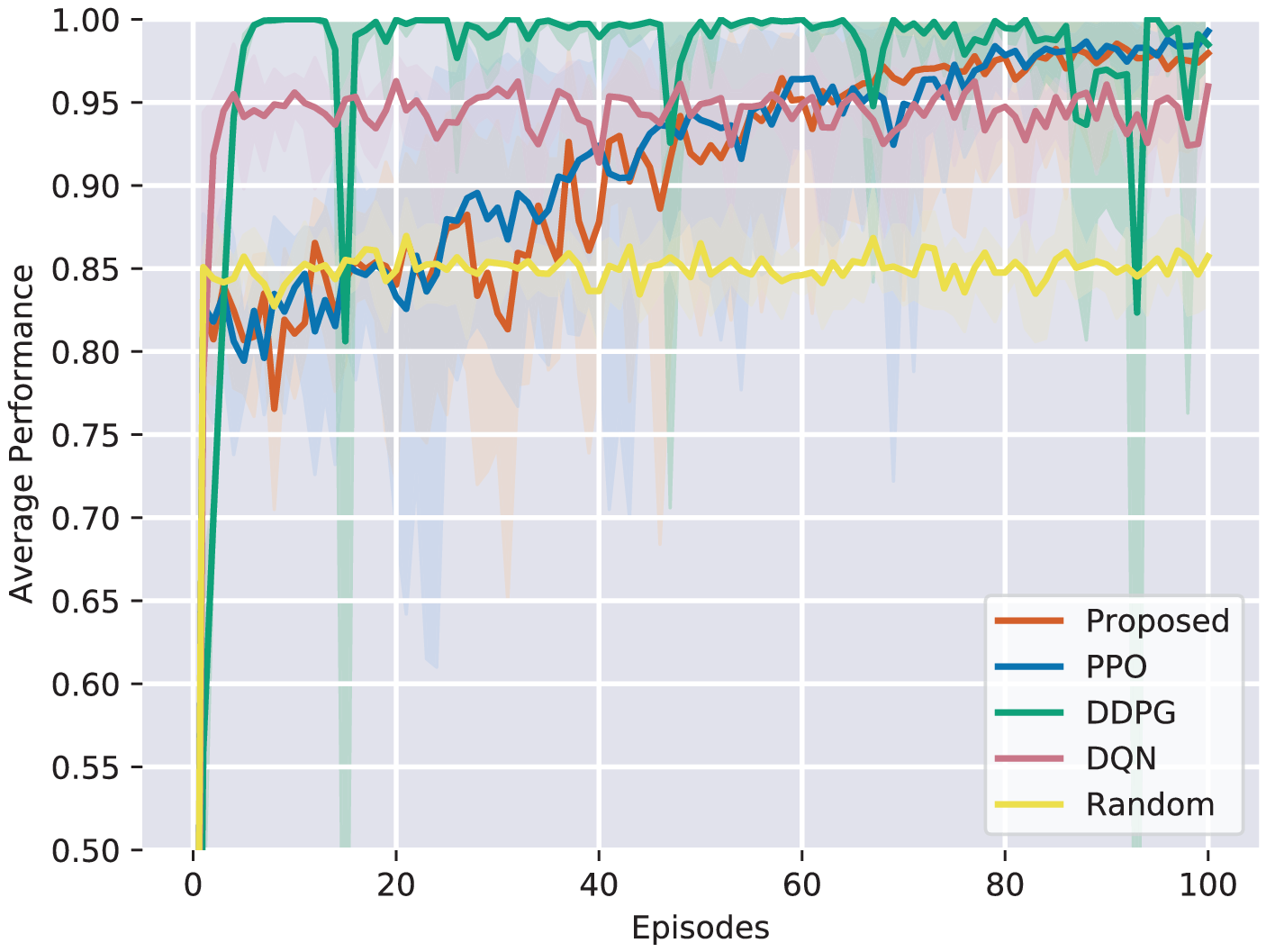}
    }
    \subfigure[ ]{
		\label{worst_train1}
		\includegraphics[scale=0.5]{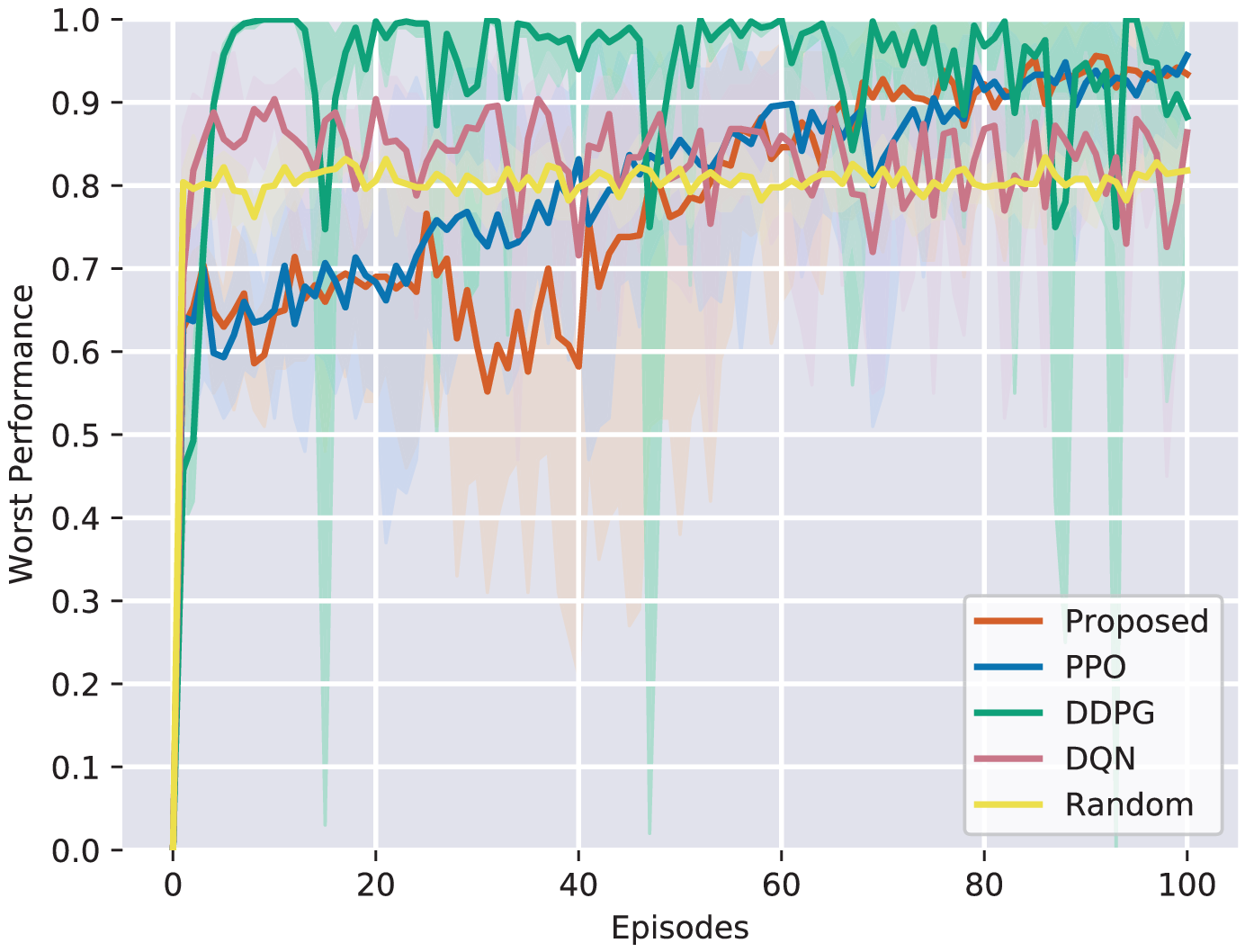}
    }

    \caption{Training performance under different methods when relay locations are unfixed.}
    \label{train1}
\end{figure*}

The training results of average performance and worst-case performance are depicted in Fig. \ref{train1}.
Note that, the average and worst-case success rate for each method are evaluated 20 times, and then mean curves and ranges are recorded.

In average performance of communication success rate, as shown in Fig. \ref{avg_train1}, we can find that our proposed method and PPO can converge to a stable value about 98\%.
The performance of PPO is similar to, or even a litter better than that of our proposed method.
It is because PPO use the full empirical samples for training, while our method screens these samples during robust design.
Considering that our method is designed to guarantee the lower bound of system performance and improve generalization capability, small gaps in performance during training process are acceptable.
In addition, DDPG and DQN have the faster training speed than our method and PPO.
However, although the best performance can be 1.0, the fluctuation of DDPG performance curve is very large, which indicates that its training stability and convergence performance are not good.
In terms of DQN method, due to its discrete power level control, the average communication success rate is about 3\% lower than other three continuous action space methods.

In worst-case performance shown in Fig. \ref{worst_train1}, our method and PPO also have stable performance, with average value being about 93\% when converging.
Similar to that in average performance evaluation, the fluctuation of DDPG and DQN is obviously larger, which also shows that the trained model is unstable.

After training, we can obtain the well-trained models equipped with dynamic relay selection and power allocation policies, which represent the actual performance of the corresponding RL algorithm.
Then, we fix the neural network parameters in these models (that is, without further training), and evaluate the robustness of each RL algorithm by taking all possible locations of each relay into consideration.

We test the average and worst-case communication success rate for algorithms used in training process in another 100 episodes.
Then, the maximum (MAX), minimum (MIN), mean (MEAN), and standard deviation (STDEV) value are recorded.
The "MAX" and "MIN" item represent the maximum and the minimum result when evaluating average/worst-case communication success rate respectively, which can also be regarded as the upper and lower bounds of the algorithm performance.
The "MEAN" item represents the average performance of RL methods in testing environments, and the "STDEV" item is used to measure the stability of them.

 \begin{table*}[htb]
        \renewcommand\arraystretch{1.4}
    	\centering
        \caption{Testing performance under different methods when relay locations are unfixed.}
		\label{test1}
        \setlength{\tabcolsep}{3mm}
        {
		\begin{tabular}{c |c |c |c |c |c |c |c |c }
        \hline
            \multirow{2}{*}{Algorithm}
            & \multicolumn{4}{c|}{Average Success Rate of RL Models} & \multicolumn{4}{c}{Worst-case Success Rate of RL Models} \\
            \cline{2-9}
            & MAX & MIN & MEAN & STDEV & MAX & MIN & MEAN & STDEV \\
        \hline
            Proposed & 0.966 & \textbf{0.856} & \textbf{0.914} & \textbf{0.0316}    & 0.915 & \textbf{0.764} & \textbf{0.839} & \textbf{0.0460} \\
            PPO & 0.955 & 0.780 & 0.899 & 0.0401    & 0.906 & 0.703 & 0.827 & 0.0475 \\
            DDPG & 0.958 & 0.005 & 0.743 & 0.3468    & 0.873 & 0 & 0.468 & 0.2983 \\
            DQN & \textbf{0.986} & 0.314 & 0.872 & 0.1484    & \textbf{0.980} & 0.146 & 0.673 & 0.2434 \\
            Random & 0.857 & 0.813 & 0.838 & /    & 0.826 & 0.749 & 0.795 & / \\
        \hline
		\end{tabular}
        }
	\end{table*}

The testing results of our first experiment are depicted in table \ref{test1}.
It can be found that, in the face of unseen environments, the lower bound performance of both average and worst-case situations of those RL algorithms without robust design is lower than that of random selection.
On the contrary, our method can still have improvements in both average and worst-case situations.
When compared with PPO, the proposed method can improve about 7.6\% and 6.1\% of the minimum value of average/worst-case success rate respectively, suggesting that our method is helpful to improve the lower bound of the system performance.
For the "MEAN" item, the average result of our method is about 1.5\% higher than that of PPO when calculating the average communication success rate, and is about 1.2\% higher than that of PPO when calculating the worst-case communication success rate, which indicates that our method can also achieve good average performance.
When compared with DDPG and DQN, their standard deviations are much larger than our method, which reflects the instability of algorithm again.
As shown in the table, although these two methods can work well in some environments, their performance can be very poor in others and it is unacceptable.
Therefore, in such changeable communication environments with unfixed relays, the proposed robust RL design can successfully improve the lower bound of worst-case performance compared to other traditional RL methods.

\begin{figure*}[hbt]
	\centering

    \vspace{-0.35cm} 
	\subfigtopskip=1pt 
	\subfigbottomskip=1pt 
	\subfigcapskip=-4pt 

    \subfigure[ ]{
		\label{avg_train2}
		\includegraphics[scale=0.5]{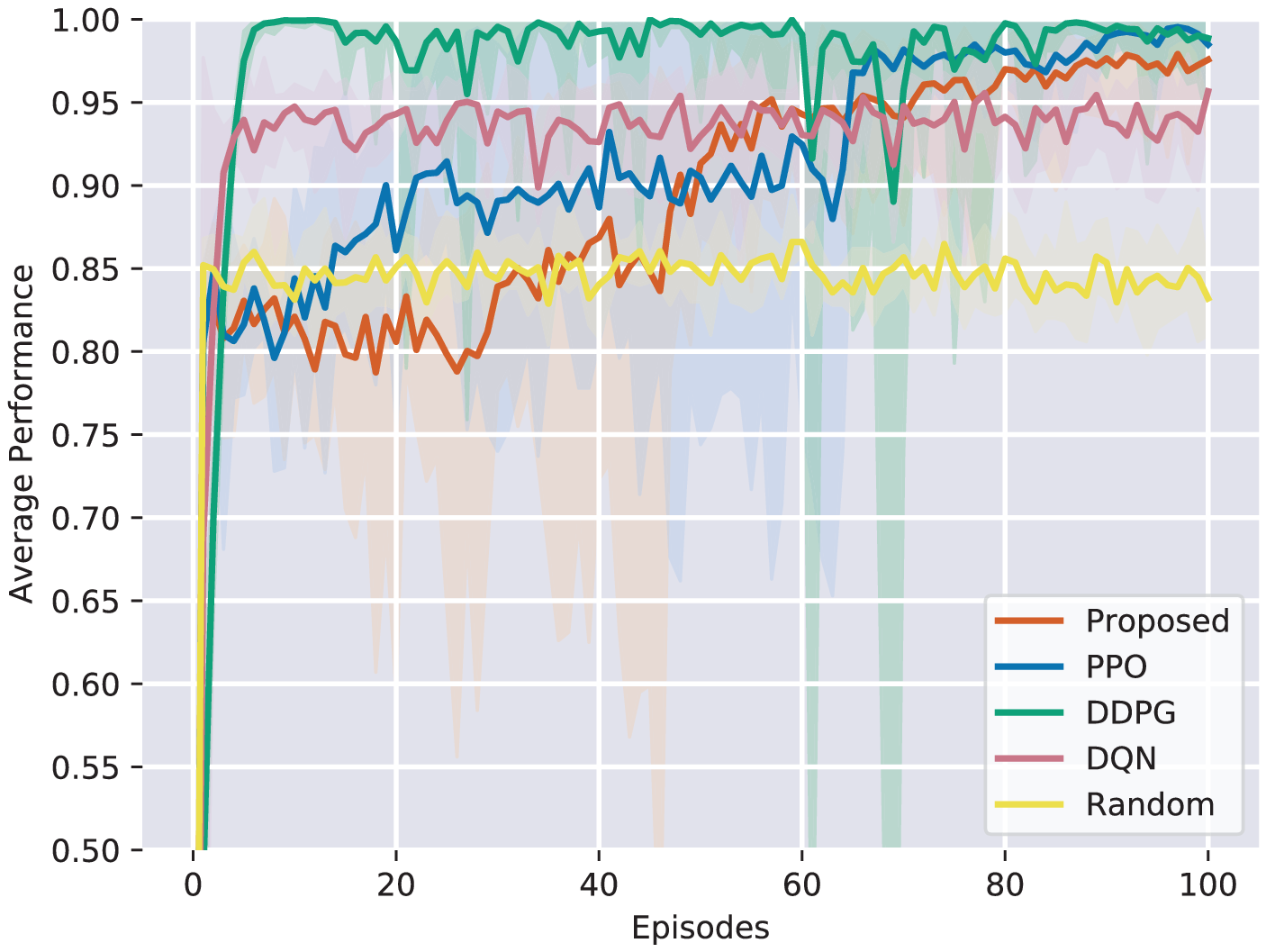}
    }
    \subfigure[ ]{
		\label{worst_train2}
		\includegraphics[scale=0.5]{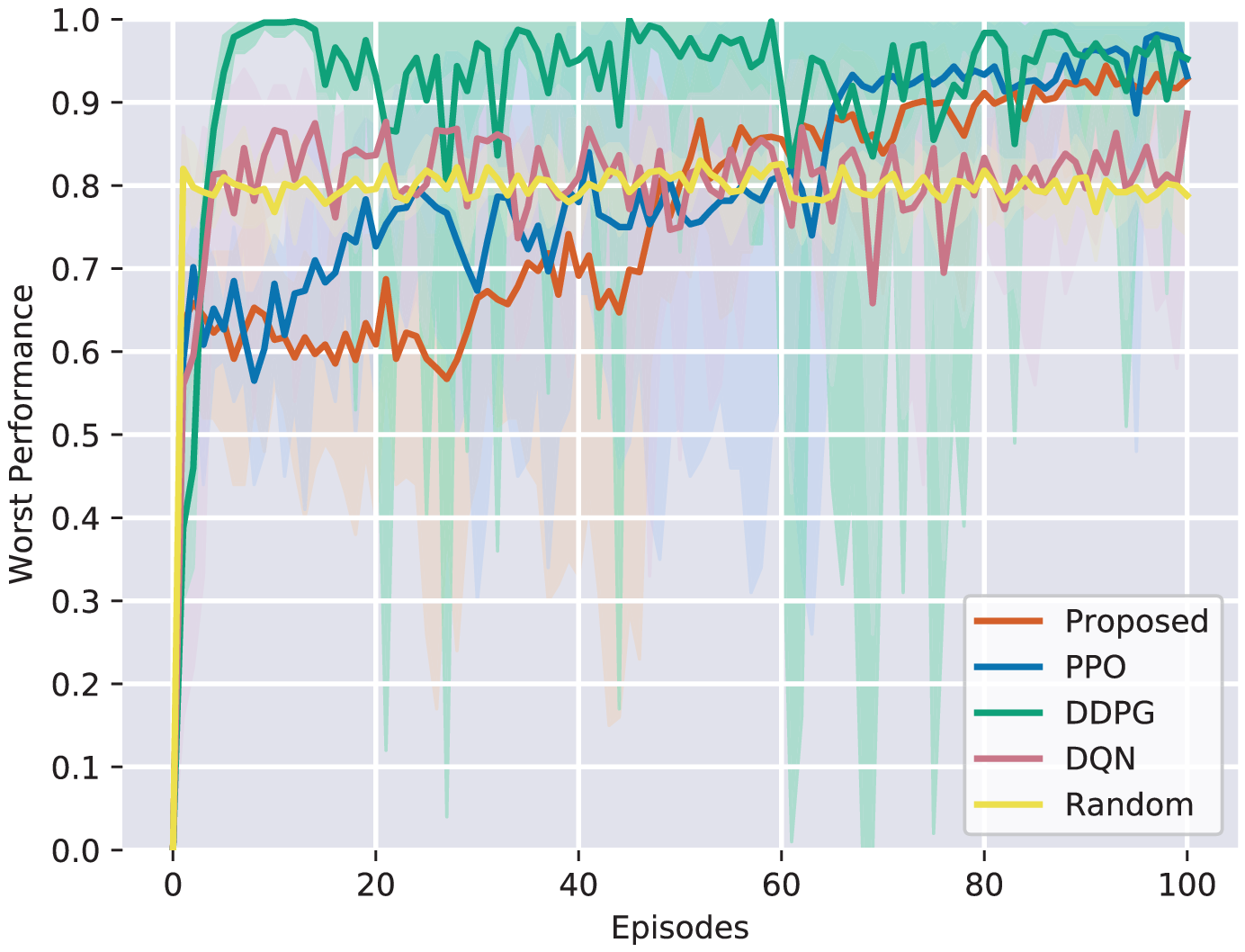}
    }

    \caption{Training performance under different methods when locations and outage thresholds of relays are unfixed.}
    \label{train2}
\end{figure*}

In the first experiment, we choose the location as the environment parameter, which can influence the transition dynamics between system states.
Further, we conduct a second experiment as supplement, where both location and threshold are used as environment parameters.
The setup is similar to the previous experiment.
We train with three of all possible outage thresholds for each relay and test with all possible thresholds, in order to evaluate the performance of different approaches in the context of changing transition dynamics and reward function.
The training results of the second experiment are shown in Fig. \ref{train2}.

In terms of average communication success rate, PPO and DDPG finally converge to a similar value, which is about 1\% higher than that of our method and 4\% higher than that of DQN.
When it comes to worst-case communication success rate, our method, PPO and DDPG have the similar performance when converging, and is about 10\% better than DQN.
Besides, the fluctuations of DDPG curve and DQN curve are obviously larger than that of ours and PPO, indicating that the training stability of our approach is better.
These conclusions are similar to those from the previous experiment.
Besides, the small average performance gap between our approach and others is acceptable because we focus primarily on improving the worst-case performance, even generalizing to unseen environments.

Similarly, we next evaluate the RL models after training process is finished.
Testing environment includes all possible locations and outage thresholds of each relay, and the results are shown in table \ref{test2}.

   \begin{table*}[h]
        \renewcommand\arraystretch{1.4}
    	\centering
        \caption{Testing performance under different methods when locations and outage thresholds of relays are unfixed.}
		\label{test2}
        \setlength{\tabcolsep}{3mm}
        {
		\begin{tabular}{c |c |c |c |c |c |c |c |c }
        \hline
            \multirow{2}{*}{Algorithm}
            & \multicolumn{4}{c|}{Average Success Rate of RL Models} & \multicolumn{4}{c}{Worst-case Success Rate of RL Models} \\
            \cline{2-9}
            & MAX & MIN & MEAN & STDEV & MAX & MIN & MEAN & STDEV \\
        \hline
            Proposed & 0.955 & \textbf{0.849} & 0.906 & \textbf{0.0232}    & 0.853 & \textbf{0.732} & \textbf{0.801} & \textbf{0.0276} \\
            PPO & 0.953 & 0.789 & 0.891 & 0.0449    & 0.866 & 0.670 & 0.780 & 0.0474  \\
            DDPG & \textbf{0.997} & 0.486 & 0.898 & 0.1254    & 0.972 & 0.024 & 0.619 & 0.3221  \\
            DQN & 0.989 & 0.659 & \textbf{0.907} & 0.0787    & \textbf{0.985} & 0.310 & 0.722 & 0.2088  \\
            Random & 0.855 & 0.814 & 0.834 & /    & 0.832 & 0.720 & 0.786 & /  \\
        \hline
		\end{tabular}
        }
	\end{table*}

It can be found that our method has the best result when evaluating the minimum value of both average success rate and worst-case success rate.
In addition, the standard deviation of our method is the smallest, indicating that the performance is more stable under different environments.
More specifically, when compared with PPO, our approach can improve about 6\% in lower bound of both average and worst-case success rate.
DDPG and DQN achieve the higher best-case value when testing in changeable environments, however, they perform poorly in terms of the lower bound performance assessment.
Our method has an improvement of about 36.3\% in minimum value of average success rate and about 70.8\% in minimum value of worst-case success rate respectively when compared with DDPG.
Similarly, in terms of lower bound performance of average success rate and worst-case success rate, our method has an improvement of about 19\% and 42.2\% respectively when compared with DQN.

In all, by combining the results in table \ref{test1} and table \ref{test2}, we can draw the conclusion that our proposed robust RL design has better worst-case performance in a changeable communication environment, and has better generalization ability for scenarios which are unseen in the training process.

\section{Conclusion}\label{sect conclusion}
In this paper, we model a two-hop DF relaying network with the uncertainty in relay locations, and study the outage probability minimization problem under total power constraints.
Without exact instantaneous CSI and prior knowledge of the communication environment, we try RL methods to solve this optimization problem.
We derive the lower bound of the worst-case performance of RL action policy, and then propose a robust RL algorithm for relay selection and power allocation.
Simulation results show that the proposed approach has better generalization ability, and the worst-case performance is improved by about 6\% compared with traditional RL methods.

\begin{appendices}
\section{Proof of Lemma 1}\label{proof_lemma1}
\textbf{Lemma 1. } \textit{Suppose two joint distributions $p_1(x,y)=p_1(x)p_1(y|x)$ and $p_2(x,y)=p_2(x)p_2(y|x)$, the total variation distance of the joint can be bounded by:}
\begin{equation}\label{lemma_1}\begin{aligned}
 D_{TV}\big(p_1(x,y)||p_2(x,y)\big) \leq \mathbb{E}_{x \sim p_1}\big[D_{TV}\big(p_1(y|x)||p_2(y|x)\big)\big] + D_{TV}\big(p_1(x)||p_2(x)\big).
\end{aligned}\end{equation}

\textbf{Proof. } For any time slot $t \in \mathbb{N}^{\ast}$, we have
\begin{equation}\begin{aligned}
&D_{TV}\big(p_1(x,y)||p_2(x,y)\big) = \frac{1}{2} \sum\limits_x \sum\limits_y | p_1(x,y)-p_2(x,y) |  \\
&= \frac{1}{2} \sum\limits_x \sum\limits_y | p_1(x)p_1(y|x) - p_2(x)p_2(y|x) |  \\
&= \frac{1}{2} \sum\limits_x \sum\limits_y | p_1(x)p_1(y|x) - p_1(x)p_2(y|x) + p_1(x)p_2(y|x) - p_2(x)p_2(y|x) |  \\
&\leq \frac{1}{2} \sum\limits_x \sum\limits_y p_1(x) |p_1(y|x)-p_2(y|x)| + \frac{1}{2} \sum\limits_x \sum\limits_y p_2(y|x) | p_1(x)-p_2(x)|  \\
&= \sum\limits_x p_1(x) \Big( \frac{1}{2}\sum\limits_y |p_1(y|x)-p_2(y|x)| \Big) + \frac{1}{2} \sum\limits_x \Big( \sum\limits_y p_2(y|x) \Big) | p_1(x)-p_2(x)|  \\
&= \mathbb{E}_{x \sim p_1} \big[D_{TV}\big(p_1(y|x)||p_2(y|x)\big) \big] + D_{TV}\big(p_1(x) || p_2(x)\big)
\end{aligned}\end{equation}
$\hfill \blacksquare$

\section{Proof of Lemma 2}\label{proof_lemma2}
\textbf{Lemma 2. } \textit{Suppose the state distributions $p_1(s_1)$ and $p_2(s_1)$ in initial time slot are the same, the total variation distance in the state marginal can be bounded by:}
\begin{equation}\begin{aligned}
 D_{TV}\big(p_1(s_t)||p_2(s_t)\big) \leq (t-1)\max_t \mathbb{E}_{s_{t-1} \sim p_1} D_{TV}\big(p_1(s_t|s_{t-1})||p_2(s_t|s_{t-1}) \big).
\end{aligned}\end{equation}

\textbf{Proof. } For any time slot $t \geq 2$ and $t \in \mathbb{N}^{\ast}$, we have
\begin{equation}\begin{aligned}
&|p_1(s_t)-p_2(s_t)| = | \sum\limits_{s_{t-1}} p_1(s_t|s_{t-1})p_1(s_{t-1}) - \sum\limits_{s_{t-1}} p_2(s_t|s_{t-1})p_2(s_{t-1}) | \\
&\leq \sum\limits_{s_{t-1}} | p_1(s_t|s_{t-1})p_1(s_{t-1}) - p_2(s_t|s_{t-1})p_2(s_{t-1}) |  \\
&= \sum\limits_{s_{t-1}} | p_1(s_t|s_{t-1})p_1(s_{t-1}) - p_2(s_t|s_{t-1})p_1(s_{t-1}) +
  p_2(s_t|s_{t-1})p_1(s_{t-1}) - p_2(s_t|s_{t-1})p_2(s_{t-1}) |  \\
&\leq \sum\limits_{s_{t-1}} p_1(s_{t-1}) |p_1(s_t|s_{t-1})-p_2(s_t|s_{t-1})| + \sum\limits_{s_{t-1}} p_2(s_t|s_{t-1})|p_1(s_{t-1})-p_2(s_{t-1})|  \\
&= \mathbb{E}_{s_{t-1} \sim p_1}|p_1(s_t|s_{t-1})-p_2(s_t|s_{t-1})| + \sum\limits_{s_{t-1}} p_2(s_t|s_{t-1})|p_1(s_{t-1})-p_2(s_{t-1})|
\end{aligned}\end{equation}
Then, by using the result above, we have
\begin{equation}\begin{aligned}
&D_{TV}\big(p_1(s_t)||p_2(s_t)\big) = \frac{1}{2} \sum\limits_{s_t}| p_1(s_t)-p_2(s_t) |  \\
&\leq \frac{1}{2} \sum\limits_{s_t} \Big[ \mathbb{E}_{s_{t-1} \sim p_1}|p_1(s_t|s_{t-1})-p_2(s_t|s_{t-1})| + \sum\limits_{s_{t-1}} p_2(s_t|s_{t-1})|p_1(s_{t-1})-p_2(s_{t-1})| \Big]  \\
&= \mathbb{E}_{s_{t-1} \sim p_1} \Big[ \frac{1}{2} \sum\limits_{s_t}|p_1(s_t|s_{t-1})-p_2(s_t|s_{t-1})| \Big]
   + \frac{1}{2}\sum\limits_{s_{t-1}} \Big( \sum\limits_{s_t} p_2(s_t|s_{t-1}) \Big) |p_1(s_{t-1})-p_2(s_{t-1})| \\
&= \mathbb{E}_{s_{t-1} \sim p_1} D_{TV}\big( p_1(s_t|s_{t-1}) || p_2(s_t|s_{t-1}) \big) + D_{TV}\big( p_1(s_{t-1}) || p_2(s_{t-1}) \big)  \\
&= \sum\limits_{i=2}^t \mathbb{E}_{s_{i-1} \sim p_1} D_{TV}\big( p_1(s_i|s_{i-1}) || p_2(s_i|s_{i-1}) \big) + D_{TV}\big( p_1(s_1) || p_2(s_1) \big) \\
&\leq (t-1)\max_t \mathbb{E}_{s_{t-1} \sim p_1} D_{TV}\big(p_1(s_t|s_{t-1})||p_2(s_t|s_{t-1})\big)
\end{aligned}\end{equation}
$\hfill \blacksquare$

\end{appendices}

\bibliographystyle{IEEEtran}
\bibliography{paper_ref}

\end{document}